\documentclass[conference]{IEEEtran}
\usepackage{flushend}
\usepackage{graphicx}
\usepackage{amssymb}
\usepackage{amsthm}
\usepackage{amsmath}
\usepackage{amsfonts}
\usepackage{subfigure}
\usepackage{lineno}
\usepackage{clrscode}
\usepackage{algorithm}
\usepackage{algpseudocode}
\usepackage{amsmath}
\newtheorem{remark}{Remark}

\ifCLASSINFOpdf
\else
\fi
\begin{document}
%
\title{Elite Bases Regression: A Real-time Algorithm for Symbolic Regression}

\author{\IEEEauthorblockN{Chen Chen\IEEEauthorrefmark{2}\IEEEauthorrefmark{3},
Changtong Luo\IEEEauthorrefmark{1}\IEEEauthorrefmark{2},
Zonglin Jiang\IEEEauthorrefmark{2}\IEEEauthorrefmark{3}}
\IEEEauthorblockA{\IEEEauthorrefmark{2}State Key Laboratory of High Temperature Gas Dynamics, Institute of Mechanics, Chinese Academy of Sciences \\ Beijing 100190, China}
\IEEEauthorblockA{\IEEEauthorrefmark{3}School of Engineering Sciences, University of Chinese Academy of Sciences, Beijing 100049, China}
\IEEEauthorrefmark{1}Email: luo@imech.ac.cn}


%


\maketitle

\begin{abstract}
Symbolic regression is an important but challenging research topic in data mining. It can detect the underlying mathematical models. Genetic programming (GP) is one of the most popular methods for symbolic regression. However, its convergence speed might be too slow for large scale problems with a large number of variables. This drawback has become a bottleneck in practical applications. In this paper, a new non-evolutionary real-time algorithm for symbolic regression, Elite Bases Regression (EBR), is proposed. EBR generates a set of candidate basis functions coded with parse-matrix in specific mapping rules. Meanwhile, a certain number of elite bases are preserved and updated iteratively according to the correlation coefficients with respect to the target model. The regression model is then spanned by the elite bases. A comparative study between EBR and a recent proposed machine learning method for symbolic regression, Fast Function eXtraction (FFX), are conducted. Numerical results indicate that EBR can solve symbolic regression problems more effectively.
\end{abstract}


%
\IEEEpeerreviewmaketitle

\section{Introduction}
\label{Section1}
Symbolic regression aims to find a mathematical model that can describe and predict a given system based on observed input-response data. It plays an increasingly important role in many engineering applications including signal processing \cite{Yang2005}, system identification \cite{Patelli2010}, industrial data analysis \cite{Luo2015}, etc. Unlike conventional linear/nonlinear regression methods that assume a linear/nonlinear trend, or require you to provide a mathematical model of a given form, symbolic regression searches an appropriate model from a space of all possible expressions $\mathcal{S}$ defined by a set of given arithmetic operations (e.g., $+$, $-$, $\times$, $\div$, etc.) and mathematical functions (e.g., $\sin$, $\cos$, $\exp$, $\ln$, etc.). Mathematically, symbolic regression finds the best combination of these operations and functions, and optimizes the model structure and coefficients simultaneously, which can be described as follows:
\begin{equation}
\label{SR}
f^*=\arg\min_{f \in \mathcal{S}}\sum_i\left\|{f(\boldsymbol{x}^{(i)})-y_i}\right\|,
\end{equation}
where $\boldsymbol{x}^{(i)}\in{\mathbb{R}^d}$, $y_i\in{\mathbb{R}}$ are numeric input-response data, and $f$ is the model function. However, given that symbolic regression is a kind of global optimization problem in data mining, there are still several difficulties in dealing with both \emph{structure optimization} and \emph{coefficient optimization} at the same time \cite{Davidson2003}. Hence, how to use a appropriate method to solve a symbolic regression problem is considered as a kaleidoscope in this research field \cite{Dong2015}.

Genetic programming (GP) \cite{Koza1992}, as a evolutionary computing (EC) technique, is one of the most popular methods for symbolic regression in recent years. Corresponding different improved versions of basic GP have also been proposed continually, for instance, linear genetic programming (LGP) \cite{Holmes1996}, gene expression programming (GEP) \cite{Ferreira2002}, parse-matrix evolution (PME) \cite{Luo2012}, etc. The core idea of GP is to apply Darwin's theory of natural evolution to the artificial world of computers and modeling. Theoretically, GP can get accurate results provided that the computation time is long enough. However, due to its stochasticity, GP is difficult to realize the real-time calculation and hard to give repeated results. In addition, the convergence speed of GP might be too slow for large scale problems with a large number of variables. Hence, GP's practical applications are limited.

To overcome these difficulties, more recently, a number of researchers have focused mainly on using non-evolutionary optimization methods to solve symbolic regression problems. McConaghy \cite{McConaghy2011} presented the first non-evolutionary algorithm based on machine learning for symbolic regression, which confined its search space to generalized linear space. Icke \& Bongard \cite{Icke2013} proposed a hybrid algorithm which combined deterministic machine learning method and conventional GP. Worm \cite{Worm2016} introduced a deterministic machine learning algorithm, Prioritized Grammar Enumeration (PGE), in his thesis, which made a large reduction to the search space. Deklel et al. \cite{Deklel2016} presented a new approach based on artificial neural networks, which could solve problems with large number of inputs and even more complex examples and applications.

Among these non-evolutionary methods, FFX is the first deterministic symbolic regression implementation. Based on generalized linear model (GLM), FFX applies a kind of machine learning method, namely pathwise regularized learning, to identify the best coefficients and bases in GLM, which is given by
\begin{equation}
\label{PRL}
\beta^*=\min \Vert {\boldsymbol y}-\boldsymbol B(x)\cdot \beta \Vert^2+\lambda_2\Vert\beta\Vert^2+\lambda_1\Vert\beta\Vert_1 ,
\end{equation}
where $\boldsymbol B(x) = \left( \phi_1(x), \phi_2(x), ..., \phi_N(x) \right)$ represents the vector of $N$ univariate and bivariate generated bases and $\beta$ is the regression parameter of GLM. $\lambda_1$ and $\lambda_2$ are set to $\lambda_1=\lambda$ and $\lambda_2=(1-\rho)\lambda$ respectively, where $\lambda$ is the regularization weight. It is reported that FFX can be an order of magnitude faster than GP, and has been successfully applied to analog circuit design and modeling \cite{Gielen2013}, the reliability analysis for analog circuit \cite{Maricau2012}, etc.

However, note that pathwise regularized learning (\ref{PRL}) needs to solve a quadratic optimization problem, which is equal to solve a large linear system. With the increase of basis number $N$, the computation cost will increase quadratically. This restricts the speed of FFX in further promote for large scale problems.

In this paper, we propose a new non-evolutionary algorithm, Elite Bases Regression (EBR), to solve symbolic regression problems. Different from FFX, most of the generated bases are discarded, and simultaneously, only elite bases are preserved and updated iteratively according to the correlation coefficients with respect to the target model. The regression model is then spanned by the elite bases. This makes EBR do not need to solve a large-scale system of linear equations. Hence, it can save computation time with little memory overhead. The performance of EBR is compared with FFX. Numerical results indicate that EBR has lower normalized mean square error (NMSE) and concise regression models than FFX's.

The presentation of this paper is organized as follows: related concepts used in EBR algorithm is introduced in Section \ref{Section2}, EBR algorithm is described in Section \ref{Section3}, and experiments and results are presented and discussed in Section \ref{Section4}. The paper is concluded in Section \ref{Section5} with remarking the future work.

\section{Related concepts used in EBR algorithm}
\label{Section2}

\subsection{Generalized linear model}
\label{Section2.1}
Generalized linear model (GLM) \cite{Nelder1972} is a generalization of classical linear regression model. GLM aims to find $f^*$ in a finite dimensional space of functions spanned by a set of given basis functions. In other words, GLM specifies a set of bases $\phi_0, \phi_1, ...,\phi_N$ from $\mathbb{R}^N$ to $\mathbb{R}$ and finds $f^*$ in the form of a linear combination of $N$ basis functions $\phi_i, i=1,2,...,N$:
\begin{equation}
\label{GLM}
f^*=\beta_0+\sum_{i=1}^N\beta_i\phi_i(x),
\end{equation}
where $\beta_i$ is the regression parameter. In Elite Basis Regression (EBR) algorithm, the regression model is spanned by a set of elite bases. The number of elite bases is denoted by $n_{presv}$.

EBR algorithm is inspired from the expansion method in mathematics. In theoretical analysis, two main methods of linear expansion, namely Taylor series and Fourier series, are very powerful tools that are widely applied to many research fields \cite{Moller1997,Deuze1989}. However, Taylor series can only be used in local expansion (the neighborhood of a certain point) with special functions, while Fourier series is utilized in periodic functions exclusively. We hope to find a \emph{global and universal} expansion strategy in practical applications. This motivates us to design such a kind of \emph{global and universal} linear approximation method for symbolic regression.

\subsection{Correlation coefficient}
\label{Section2.2}
Correlation coefficient aims to measure linear relationship between two vectors, and has a wide application scope in statistical analysis. Suppose a function with $n$ continuous variables
\begin{equation}
\label{func2.2}
f(x)=f\left(x_1, x_2,...,x_n\right),x_i\in [a_i,b_i],i=1,2,...,n.
\end{equation}
For each variable $x_i$, a column vector $\boldsymbol{x_i}$ is defined after a set of random sample points in $[a_i,b_i]$ are generated. That is $X_i=\boldsymbol{x_i}=\left(x_i^{(1)},x_i^{(2)},...,x_i^{(m)}\right)^{\text{T}}$, where $m$ is the number of sample points. Thus, we can get a column vector with $m$ components respect to the initial function (\ref{func2.2})
\begin{equation}
\label{corrcoeff}
\begin{aligned}
  F\left( X \right) &= F\left( {\left( {\begin{array}{*{20}{c}}
  {x_1^{\left( 1 \right)}} \\ 
  {x_1^{\left( 2 \right)}} \\ 
   \vdots  \\ 
  {x_1^{\left( m \right)}} 
\end{array}} \right),\left( {\begin{array}{*{20}{c}}
  {x_2^{\left( 1 \right)}} \\ 
  {x_2^{\left( 2 \right)}} \\ 
   \vdots  \\ 
  {x_2^{\left( m \right)}} 
\end{array}} \right), \ldots ,\left( {\begin{array}{*{20}{c}}
  {x_n^{\left( 1 \right)}} \\ 
  {x_n^{\left( 2 \right)}} \\ 
   \vdots  \\ 
  {x_n^{\left( m \right)}} 
\end{array}} \right)} \right) \\ 
   &= \left( {\begin{array}{*{20}{c}}
  {f\left( {x_1^{\left( 1 \right)},x_2^{\left( 1 \right)}, \ldots ,x_n^{\left( 1 \right)}} \right)} \\ 
  {f\left( {x_1^{\left( 2 \right)},x_2^{\left( 2 \right)}, \ldots ,x_n^{\left( 2 \right)}} \right)} \\ 
   \vdots  \\ 
  {f\left( {x_1^{\left( m \right)},x_2^{\left( m \right)}, \ldots ,x_n^{\left( m \right)}} \right)} 
\end{array}} \right) = \left( {\begin{array}{*{20}{c}}
  {{f^{\left( 1 \right)}}} \\ 
  {{f^{\left( 2 \right)}}} \\ 
   \vdots  \\ 
  {{f^{\left( m \right)}}} 
\end{array}} \right) \\ 
\end{aligned}
\end{equation}

From the above discussion, for an $n$-dimensional problem in EBR, the target model $f\left(x_1, x_2,...,x_n\right)$ and a certain generated basis $\phi\left(x_1, x_2,...,x_n\right)$ can be regarded as two column vectors of $m$ components after sampling, namely $Y=F\left( X \right)={\left( {{f^{(1)}},{f^{(2)}}, \cdots ,{f^{(m)}}} \right)^{\text{T}}}$ and $\xi_{\Phi}=\Phi\left( X \right)={\left( {{\phi^{(1)}},{\phi^{(2)}}, \cdots ,{\phi^{(m)}}} \right)^{\text{T}}}$. The correlation coefficient of these two vectors, $Y$ and $\xi_{\Phi}$, can be expressed as
\begin{equation}
\label{corrcoeff}
\rho_{\xi_{\Phi},Y}=\frac{{\text{Cov}} \left( \xi_{\Phi},Y \right) }{\sqrt{D(\xi_{\Phi})}\sqrt{D(Y)}},
\end{equation}
where the operator $D$ represents variance.

Note that the basis functions might be nonlinear, but the ensemble process of the GLM is still linear. Therefore, correlation test is effective in EBR. That is, if $|\rho_{\xi_{\Phi},Y}|$ is close to 1, $Y$ and $\xi_{\Phi}$ are closely related. Particularly, when $ |\rho_{\xi_{\Phi},Y}|=1 $, $Y$ and $\xi_{\Phi}$ are linearly correlated in probability one. 

\begin{figure}
\centering
\subfigure[The test function.]{
\includegraphics[width=0.6\linewidth] {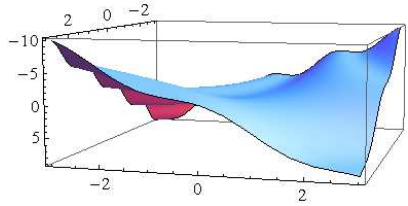}
\label{case21a}
}
\subfigure[The basis function.]{
\includegraphics[width=0.6\linewidth] {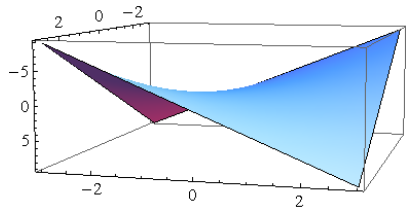}
\label{case22b}
}
\caption{An example of correlation analysis applied to 2D nonlinear function.}
\label{cof}
\end{figure}

In fact, once we regard two nonlinear functions as two column vectors after sampling, it still makes sense that correlation coefficient can be used for correlation analysis of nonlinear function. To give an illustrative example, consider a two-dimensional test function
\begin{equation}
\label{egtestfunc}
f({x_1},{x_2}) = {x_1}{x_2} + \sin \left( {\left( {{x_1} - 1} \right)\left( {{x_2} - 1} \right)} \right),
\end{equation}
and a certain basis function
\begin{equation}
\label{egbasis}
\phi(x_1,x_2)=x_1x_2,
\end{equation}
where $x\in [-3,3]^2$. To enhance the stability of the test, the distribution of sample points in $[-3,3]^2$ should be as uniform as possible by using controlled sampling method of Latin hypercube design \cite{Beachkofski2002}. Then, a correlation coefficient of the two vector functions $Y$ and $\xi_{\Phi}$ could be obtained, which is $|\rho_{\xi_{\Phi},Y}|=0.9838$. This means $Y$ and $\xi_{\Phi}$ are closely related. Note that, as is shown in Fig. \ref{cof}, the simple basis function (\ref{egbasis}) successfully `sketch out' the landscape of test function (\ref{egtestfunc}).

\subsection{Parse-matrix encoding scheme}
\label{Section2.3}
The generation of basis functions is a crucial step in EBR. However, multiple nested-loop used to enumerate the possible bases in FFX is not easy to extend. On the other hand, the complexity of bases is implemented with the if-else statement, which makes FFX difficult to control the complexity and limits its ability of modeling highly nonlinear target function. Hence, parse-matrix encoding scheme becomes a good candidate for bases generation engine of EBR.

Parse-matrix encoding scheme was initially provided for parse-matrix evolution (PME) \cite{Luo2012}, a special version of GP. PME use a two-dimensional matrix with integer entries to express a chromosome (individual), which can carry more information than conventional chromosome representations \cite{Goldberg1996, ONeil2003}. The matrix representation makes PME easy to control the complexity and simple to program.

\begin{figure}
\centering
\includegraphics[width=0.9\linewidth]{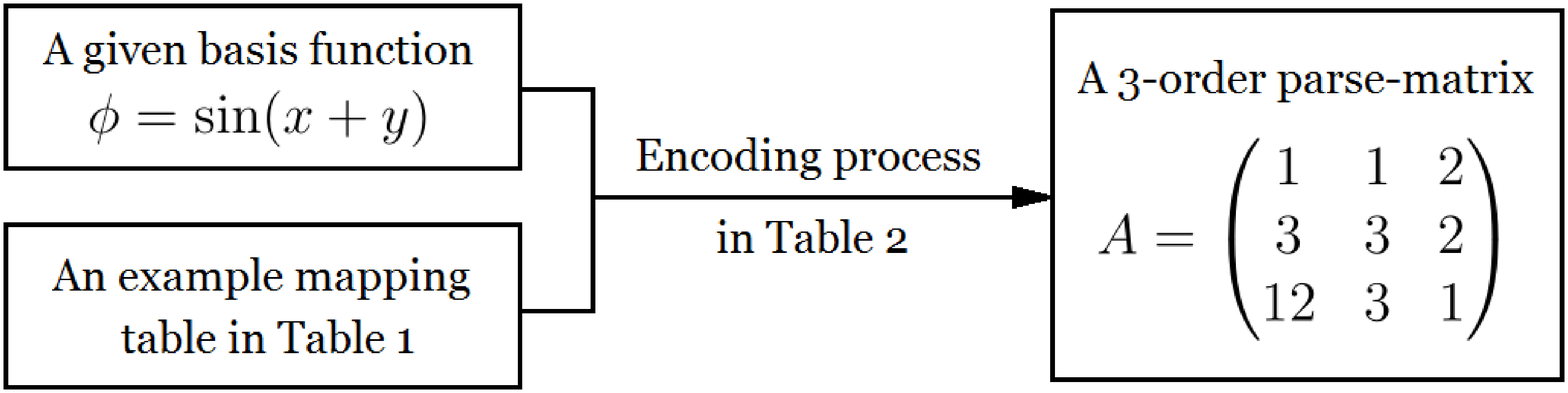}
\caption{An example of parse-matrix encoding process of EBR.}
\label{fig:BasisEncodingProcess}
\end{figure}

In EBR, basis functions are coded with parse-matrix encoding scheme. This process could be sketched in Fig. \ref{fig:BasisEncodingProcess}. Suppose an example mapping table defined as Table \ref{Table_map_basis}. Then, a given basis function $ \phi=\sin(x+y)$ can be produced in the steps listed in Table \ref{Encod_exp}. According to the mapping table (see Table \ref{Table_map_basis}) and the encoding steps (see Table \ref{Encod_exp}), the basis function $ \phi=\sin(x+y)$ can be described by a parse-matrix of order $3\times3$ as follows: 

\begin{equation}
\label{Parse_matrix}
A=\begin{pmatrix}
1 & 1 & 2 \\
3 & 3 & 2 \\
12 & 3 & 1
\end{pmatrix}
\end{equation}

\begin{table*}
\scriptsize
\centering
\caption{An example mapping table for a basis function.}
\label{Table_map_basis}
\begin{tabular}{llllllllllllll}
\hline
$a._1$   & 1  & 2  & 3 & 4 & 5 & 6 & \multicolumn{1}{c}{7} & 8   & 9    & 10  & 11  & 12  & 13  \\
$T$    & $s_1$ & $s_2$ & $+$ & $-$ & $*$ & $/$ & $\sqrt{}$                  & $s_1^2$ & $1/s_1$ & $log$ & $exp$ & $sin$ & $cos$ \\
$a._2,a._3$ & 1  & 2  & 3 &   &   &   &                       &     &      &     &     &     &     \\
$expr$ & $x_1$ & $x_2$ & $f$ &   &   &   &                       &     &      &     &     &     &     \\ \hline
\end{tabular}
\end{table*}

\begin{table}[h]
\scriptsize
\centering
\caption{Encoding steps of the basis function $\sin(x+y)$.}
\label{Encod_exp}
\begin{tabular}{lllll}
\hline
Step & $T$   & $s_1$ & $s_2$ & Update       \\ \hline
1    & $x_1$  & $x_1$ & $x_2$ & $f=x_1$         \\
2    & $+$   & $f$  & $x_2$ & $f=x_1+x_2$      \\
3    & $sin$ & $f$  & $x_1$ & $f=sin(x_1+x_2)$ \\ \hline
\end{tabular}
\end{table}

We can see that the encoding of parse-matrix is a natural and easy process. Note that the second and the third columns $a._2,a._3$ are used to control the dimension of a given problem. The parse-matrix encoding scheme ensures the generated candidate basis functions can cover all possible bases, according to the mapping rules in Table \ref{Table_map_basis}. In the following section, we will use this table to do our numerical experiments (see Section \ref{Section4}).

\section{Elite Basis Regression algorithm}
\label{Section3}
The new algorithm is a kind of deterministic linear approximation method. It does not rely on other GP method. To illustrate our proposed algorithm more clearly, in this section, EBR is introduced by two main parts, which are \emph{generation and preservation of the bases} in Section \ref{Section3.1} and \emph{ensemble and evaluation of the model} in Section \ref{Section3.2}. Then, we discuss the method of complexity control in EBR in Section \ref{Section3.3}. Finally, the whole procedure of EBR is introduced in Section \ref{Section3.4}.

\subsection{Generation and preservation of bases}
\label{Section3.1}
In EBR, the basis function is coded with parse-matrix encoding scheme in specific mapping rules (refer to Table \ref{Table_map_basis}), which has been discussed in Section \ref{Section2.3}. Enumeration method is used to generate a set of candidate basis functions. This process ensures the generated bases can cover all possibilities in given arithmetic operations (e.g., $+$, $-$, $\times$, $\div$, etc.) and mathematical functions (e.g., $\sin$, $\cos$, $\exp$, $\ln$, etc.). Simultaneously, $n_{presv}$ elite bases are preserved and updated iteratively according to the correlation coefficients with respect to the given target model.

\begin{remark}
The number of preserved elite bases $n_{presv}$ determines the most computation costs of EBR and the complexity of the regression model. A more detailed discussion of this control parameter will be conducted in Section \ref{Section4.3}.
\end{remark}

\begin{remark}
If the the correlation coefficients of two generated bases with respect to the target model, namely $\rho_{\xi_\Phi,Y}$ and $\rho_{\xi_\Phi,Y}^*$, are very close to zero (e.g., $  \left| {| {{\rho _{{\xi _\Phi },Y}}} | - | {\rho _{{\xi _\Phi },Y}^*} |} \right| <10^{-7}$), one basis function of them will be discarded.
\end{remark}

\subsection{Ensemble and evaluation of the model}
\label{Section3.2}
The regression model is established by GLM. Note that the number of bases participated in computation is $n_{presv}$, not all generated candidate bases. This makes EBR can realize real-time computation. We take normalized mean square error (NMSE) as test error in the numerical study, which is used to evaluate the regression model. The NMSE is defined by Eq. (\ref{NMSE}):

\begin{equation}
\label{NMSE}
\text{NMSE} \left( f,f^*\right) = \frac{\Vert f-f^* \Vert _2^2}{\Vert f \Vert _2^2},
\end{equation}
where the $f$ and $f^*$ are the target model and regression model, respectively.

\subsection{Complexity control}
\label{Section3.3}
In EBR, complexity control mainly includes two parts, namely the \emph{inner control} and \emph{outer control}.

\begin{figure}
\centering
\includegraphics[width=0.9\linewidth]{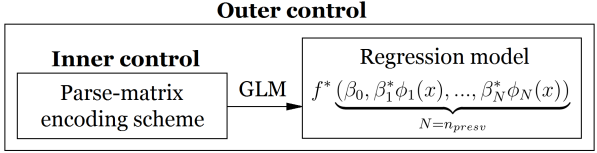}
\caption{Complexity control of EBR algorithm.}
\label{CompControl}
\end{figure}

Recall from the parse-matrix encoding scheme in Section \ref{Section2.3} that, the \emph{inner control} aims to determine the complexity of a basis function and the dimension of a given symbolic regression problem, by controlling the rows of the parse-matrix (\ref{Parse_matrix}) and the the second and the third columns $a._2,a._3$, respectively. More precisely, for a muti-dimensional problem, the entries $a_{ij}$ are bounded integers according to the mapping rules in Table \ref{Table_map_basis}, namely the parse-matrix entries $a._1\in \left\{ 1, 2, 3, ..., 13 \right\}$ and $a._j\in \left\{ 1, 2, 3, ..., d+1 \right\} (j=2,3)$, where $d$ is the dimension of the target model. The \emph{outer control} in EBR focuses on the overall complexity of the regression model, and is controlled by the prespecified $n_{presv}$. A detailed discussion of this control parameter is in Section \ref{Section4.3}. The complexity control makes EBR algorithm have a good flexibility. The relations between the \emph{inner control} and \emph{outer control} are shown in Fig. \ref{CompControl}.

\subsection{Procedure}
\label{Section3.4}
\begin{figure}
\centering
\includegraphics[width=0.9\linewidth]{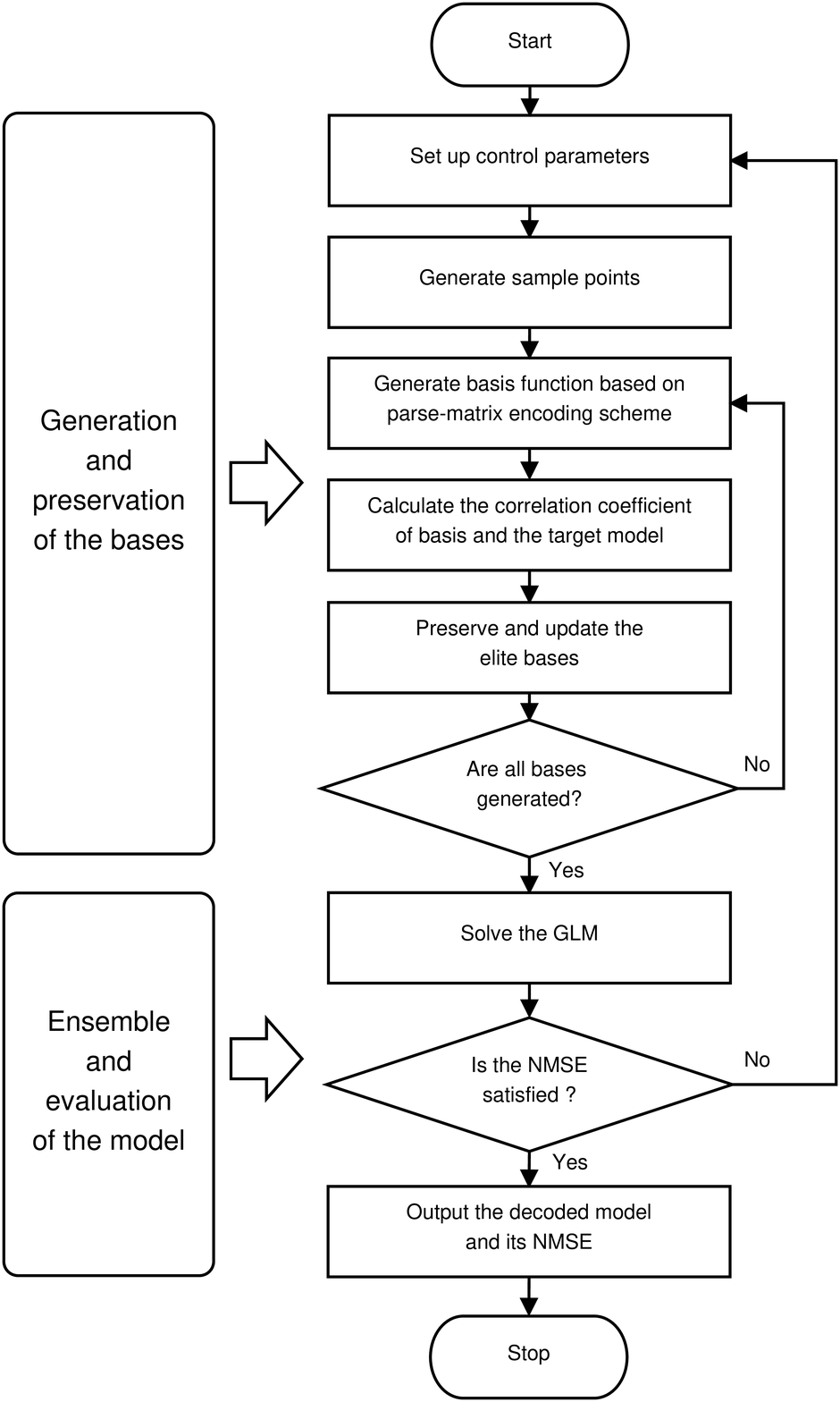}
\caption{The flow chart of EBR algorithm.}
\label{flowchart}
\end{figure}

Up to this point in our discussion, we have a general understanding of EBR algorithm. The main steps of EBR is also given in the flow-chart of EBR algorithm in Fig. \ref{flowchart}. The procedure of EBR could be described as follows.
\begin{description}
\item {\emph{Procedure of EBR:}}
\item {\emph{Step 1.}}
Initialize: Input the number of basis functions needed to be preserved $n_{presv}$, the sampling range $[a,b]$ and the test function $f$.

\item {\emph{Step 2.}}
Generate candidate basis: An enumeration method is used to generate a candidate basis function $\phi_i$ coded with parse-matrix encoding schemes.

\item {\emph{Step 3.}}
Evaluate and preserve:

(3.1) Evaluate: Evaluate each generated candidate basis by its correlation with respect to target model.

(3.2) Preserve: Preserve the basis with higher correlation with respect to target model and update the elite bases.

\item {\emph{Step 4.}}
Repeat step 2 and 3 until all possible bases are evaluated. The preserved $n_{presv}$ functions form a set of elite bases for GLM.

\item {\emph{Step 5.}}
Model: Solving the GLM (\ref{GLM}) to get regression model based on the set of elite bases. Output the decoded model and its test error (NMSE).
\end{description}

\section{Numerical results and discussion}
\label{Section4}
In order to test the performance of our proposed algorithm, several numerical experiments of classical symbolic regression problems are conducted. These problems, given in Table \ref{test_table_1}, \ref{test_table_3} and \ref{test_table_4}, are mostly taken from references \cite{Luo2012,Dervis2012}. The results are compared with machine learning algorithm Fast Function eXtraction (FFX) \cite{McConaghy2011}. NMSE is used as the test error, which is governed by Eq. (\ref{NMSE}).

The test problems are partitioned into two groups: exact fitting problems (Section \ref{Section4.1}) and linear approximation fitting problems (Section \ref{Section4.2}), to test EBR's capabilities of \emph{structure optimization} and \emph{coefficient optimization}, respectively. Additionally, we give a discussion on control parameter ($n_{presv}$) of EBR in section \ref{Section4.3}. To enhance the stability of the EBR, the distribution of sample points in should be as uniform as possible. Therefore, controlled sampling method, Latin hypercube sampling \cite{Beachkofski2002} and orthogonal sampling \cite{Steinberg2006} are preferred to generate training points (sample points).

\subsection{Exact fitting problems}
\label{Section4.1}
In this test group (Case 1-16, see Table \ref{test_table_1}), all of cases have exact fitting models, which recover the target models. In other words, these cases are chosen to test the ability of function \emph{structure optimization} for EBR, which is one of the most concerned issues in symbolic regression.

The full names of the notations in Table \ref{test_table_1} are the dimension of modeled system (Dim), the target model (Target model), the domain of the target model (Domain), the number of training points (No. samples), total bases generated by EBR (Total bases of EBR), the number of bases of EBR in final regression model (No. bases of EBR), the number of bases of FFX in final regression model (No. bases of FFX), the test error of FFX in final regression model (Test error(\%) of FFX).

\begin{table*}
\scriptsize
\centering
\caption{Test models and performance of EBR and FFX (Case 1-16).}
\label{test_table_1}
\begin{tabular}{lllllllll}
\hline
No. & Dim & Target model             & Domain     & \begin{tabular}[c]{@{}l@{}}No.\\ samples\end{tabular} & \begin{tabular}[c]{@{}l@{}}Total\\ bases\\ of EBR\end{tabular} & \begin{tabular}[c]{@{}l@{}}No.\\ bases\\ of EBR\end{tabular} & \begin{tabular}[c]{@{}l@{}}No.\\ bases\\ of FFX\end{tabular} & \begin{tabular}[c]{@{}l@{}}Test\\ error(\%)\\ of FFX\end{tabular} \\ \hline
1   & 1   & $\sqrt{x}$               & $[1,3]$    & $30$                                                  & 7488                                                           & \textbf{1}                                                   & 5                                                            & 0.869                                                             \\
2   & 1   & $x^2-\sin x$             & $[1,3]$    & $30$                                                  & 7493                                                           & \textbf{2}                                                   & 5                                                            & 0.988                                                             \\
3   & 1   & $\cos x^2-x$             & $[-3,3]$   & $30$                                                  & 5510                                                           & \textbf{1}                                                   & 8                                                            & 6.19                                                              \\
4   & 1   & $\sin x+2x$              & $[-3,3]$   & $30$                                                  & 5481                                                           & \textbf{1}                                                   & 4                                                            & 0.672                                                             \\
5   & 1   & $x^4+x^3+x^2+x$          & $[-3,3]$   & $30$                                                  & 5512                                                           & \textbf{4}                                                   & 8                                                            & 2.08                                                              \\
6   & 1   & $\sin x_1^2* \cos x_1-1$ & $[-3,3]$   & $30$                                                  & 5511                                                           & \textbf{2}                                                   & 9                                                            & 16.9                                                              \\
7   & 2   & $x_1^{x_2}$              & $[1,3]^2$  & $30^2$                                                & 7499                                                           & \textbf{1}                                                   & 8                                                            & 0.991                                                             \\
8   & 2   & $\ln(x_1+x_2)$           & $[1,3]^2$  & $30^2$                                                & 7489                                                           & \textbf{1}                                                   & 8                                                            & 0.851                                                             \\
9   & 2   & $x_1^2+x_1-x_2$          & $[-3,3]^2$ & $30^2$                                                & 5507                                                           & \textbf{4}                                                   & 8                                                            & 0.986                                                             \\
10  & 2   & $x_1+2x_2$               & $[-3,3]^2$ & $30^2$                                                & 5505                                                           & \textbf{2}                                                   & 2                                                            & 0.968                                                             \\
11  & 2   & $\sin(x_1^2-x_2)$        & $[-3,3]^2$ & $30^2$                                                & 5509                                                           & \textbf{1}                                                   & 9                                                            & 28.0                                                              \\
12  & 2   & $x_1-e^{x_1+x_2}$        & $[-3,3]^2$ & $30^2$                                                & 5489                                                           & \textbf{1}                                                   & 10                                                           & 1.00                                                              \\
13  & 2   & $(x_1+x_2)/x_2$          & $[-3,3]^2$ & $30^2$                                                & 5493                                                           & \textbf{2}                                                   & 2                                                            & 7.42                                                              \\
14  & 2   & $6\sin x_1 \cos x_2$     & $[-3,3]^2$ & $30^2$                                                & 5515                                                           & \textbf{2}                                                   & 9                                                            & 25.6                                                              \\
15  & 3   & $x_1+x_2+x_3$            & $[-3,3]^3$ & $10^3$                                                & 17163                                                          & \textbf{3}                                                   & 11                                                           & 0.987                                                             \\
16  & 3   & $x_1x_2+x_2x_3$          & $[-3,3]^3$ & $10^3$                                                & 17139                                                          & \textbf{4}                                                   & 2                                                            & 0.99                                                              \\ \hline
\end{tabular}
\end{table*}

\subsubsection{Control parameter setting}
\label{Section4.1.1}
Here, we set $n_{presv}=35$ for 1D and 2D cases, $n_{presv}=200$ for 3D cases, where $n_{presv}$ is the prespecified parameter in step 1 of EBR (see Section \ref{Section3.4}). The region is chosen as $[-3,3]$,$[-3,3]^2$ and $[-3,3]^3$ for one-dimensional (1D), 2D and 3D problems, respectively. If there is a square-root function or logarithmic function in our target model, the left endpoint of the interval is replaced to 1. The number of training point is set up to $30$, $30^2$ and $30^3$ for 1D, 2D and 3D problems, respectively. The order of the parse-matrix is fixed to 3.

\subsubsection{Numerical results}
\label{Section4.1.2}
Table \ref{test_table_1} shows the test models and performance of EBR and corresponding results of FFX. Numerical results (regression models) are listed in Table \ref{test_table_2}.

\begin{table*}
\scriptsize
\centering
\caption{Exact fitting results of EBR.}
\label{test_table_2}
\begin{tabular}{ll}
\hline
No. & Regression model                                                                     \\ \hline
1   & $f^*=0.7071*\sqrt{x+x}$                                                              \\
2   & $f^*=(-1)*(\sin x-x)+(x^2-x)$                                                        \\
3   & $f^*=\cos x^2-x$                                                                     \\
4   & $f^*=\sin x+x+x$                                                                     \\
5   & $f^*=-0.5*(xe^x-x)+0.5*(x^4+x)+0.5*(x^2+xe^x)+0.5*(x^2+x)^2$                         \\
6   & $f^*=(-1)+0.5*\sin (x_1^2+x_1)+0.5*\sin (x_1^2-x_1)$                                 \\
7   & $f^*=e^{x_2*\ln x_1}$                                                                \\
8   & $f^*=0.5*\ln (x_1+x_2)^2$                                                            \\
9   & $f^*=0.5*(x_1^2-x_2)+0.5*(x_1^2+x_1)+0.5*(e^{x_1}-2x_2)-0.5*(e^{x_1}-x_2-x_1)$       \\
10  & $f^*=x_1+x_2+x_2$                                                                    \\
11  & $f^*=\sin (x_1^2-x_2)$                                                               \\
12  & $f^*=(-1)*(e^{x_1+x_2}-x_1)$                                                         \\
13  & $f^*=0.2*(2x_1-x_2)/x_2+0.6*(2x_2+x_1)/x_2$                                          \\
14  & $f^*=3*\sin (x_1+x_2)+3*\sin (x_1-x_2)$                                              \\
15  & $f^*=0.5*(x_1+x_2)+0.25*(2x_2+2x_3)+0.25*(2x_1+2x_3)$                                \\
16  & $f^*=(-0.25)*(x_1^2-2x_1x_2)+0.25*(x_1^2+2x_1x_2)+0.25*(x_2^2)+0.25*(x_3^2+2x_2x_3)$ \\ \hline
\end{tabular}
\end{table*}

\subsubsection{Discussion}
\label{Section4.1.3}
The computation results from Table \ref{test_table_1} show that EBR can recover the target models for all these test problems (Case 1-16). In Case 5, 6, 9 and 13-16, although EBR does not find the bases involved in the corresponding target models, EBR might give the regression models in form of identities. Particularly, in Case 6 and 16, namely $f=\sin x_1^2\cos x_1-1$ and $f=6\sin x_1 \cos x_2$, note that EBR can reduce a product term to summation of trigonometric function. 

Moreover, almost all of the number of the bases in final results of EBR is far less than FFX, while the results are much better than FFX. This is because that FFX does not cover a larger operation and function space. Some cases of test errors of FFX are extremely large (namely Case 6, 11, 13 and 14), which shows that FFX is poor at providing a symbolic regression model in highly nonlinear function. From all of the results above, we can draw a conclusion that the EBR has a good capability of \emph{structure optimization} in symbolic regression problems.

\subsection{Linear approximation fitting problems}
\label{Section4.2}
As we know, practical engineering applications of symbolic regression are generally complex, so whether an algorithm can give an approximation fitting model becomes very important. The purpose of this test group is to show EBR's capability of providing an linear approximation regression model. This can be regraded as the ability of \emph{coefficient optimization}.

\begin{table*}
\scriptsize
\centering
\caption{Test models and performance of EBR and FFX (Case 17-24).}
\label{test_table_3}
\begin{tabular}{llllllll}
\hline
No. & Dim & Target model                                & Domain     & \begin{tabular}[c]{@{}l@{}}No.\\ bases\\ of EBR\end{tabular} & \begin{tabular}[c]{@{}l@{}}Test\\ error(\%)\\ of EBR\end{tabular} & \begin{tabular}[c]{@{}l@{}}No.\\ bases\\ of FFX\end{tabular} & \begin{tabular}[c]{@{}l@{}}Test\\ error(\%)\\ of FFX\end{tabular} \\ \hline
17  & 1   & $0.3x\sin (2\pi x)$                         & $[-3,3]$   & \textbf{7}                                                   & \textbf{4.37}                                                     & 10                                                           & 21.09                                                             \\
18  & 1   & $\ln (x+1)+\ln (x^2+1)$                     & $[1,3]$    & \textbf{4}                                                   & \textbf{6.58e-10}                                                 & 5                                                            & 0.967                                                             \\
19  & 1   & $x^5+x^4+x^3+x^2+x$                         & $[-3,3]$   & \textbf{7}                                                   & \textbf{1.45e-7}                                                  & 6                                                            & 1.91                                                              \\
20  & 1   & $x^6+x^5+x^4+x^3+x^2+x$                     & $[-3,3]$   & \textbf{11}                                                   & \textbf{4.97e-4}                                                  & 7                                                            & 2.08                                                              \\
21  & 2   & $\ln (x_1+x_2)+\sin (x_1+x_2)$              & $[1,3]^2$  & \textbf{10}                                                  & \textbf{1.19e-2}                                                  & 8                                                            & 16.25                                                             \\
22  & 2   & $x_1^4-x_1^3+x_2^2/2-x_2$                   & $[-3,3]^2$ & \textbf{5}                                                   & \textbf{5.87e-2}                                                  & 16                                                           & 3.47                                                              \\
23  & 2   & $x_1^3/5+x_2^3/2-x_2-x_1$                   & $[-3,3]^2$ & \textbf{7}                                                   & \textbf{0.26}                                                     & 12                                                           & 0.991                                                             \\
24  & 2   & $x_1x_2+\sin \left( (x_1-1)(x_2-1) \right)$ & $[-3,3]^2$ & \textbf{4}                                                   & \textbf{3.69}                                                     & 14                                                           & 4.18                                                              \\ \hline
\end{tabular}
\end{table*}

\subsubsection{Control parameter setting}
\label{Section4.2.1}
Similar to the first test group, we set $n_{presv}=35$ to all cases. The number of training point is set up to $30$ and $30^2$ for 1D and 2D problems, respectively. The order of the parse-matrix is fixed to 3. 

\subsubsection{Numerical results}
\label{Section4.2.2}
An overview numerical results are listed in Table \ref{test_table_3}.

\subsubsection{Discussion}
\label{Section4.2.3}
In this test group, 8 cases show the performance of EBR for its linear approximation fitting, which could be regarded as a capability of \emph{coefficient optimization}. To recap briefly for Section \ref{Section2.1}, EBR is deemed to a linear approximation method based on GLM. We hope to find a \emph{global and universal} expansion strategy, different from Taylor series and Fourier series. 

Note that the regression models of EBR are closer to the target model. For most cases, EBR performs better than FFX for its succinct regression models (less number of bases), especially for highly nonlinear target models (Case 17, 21 and 24). Meanwhile, the comparison of NMSEs in Table \ref{test_table_2} indicates that EBR has much lower NMSE at all cases. EBR exhibits reasonable accuracy, which indicate that the proposed algorithm EBR can fit the target functions in forms of polynomial functions, trigonometric, logarithmic and bivariate functions. Good performances for modeling target functions show the potential of EBR to be applied in practical applications.

\subsection{Study on control parameters}
\label{Section4.3}
The paramount control parameter in EBR is the $n_{presv}$. In this part, the value of $n_{presv}$ to be set is different from the previous test groups, in order to study its influence on the regression model. The results of this part (Case 25-28) is given in Table \ref{test_table_4}. Note that the target models given in Table \ref{test_table_4} are all highly nonlinear functions in 1D and 2D. 

Using the given control parameter $n_{presv}=35$, EBR failed to get the exact fitting models or the approximate models with NMSE $\leq 5\%$ in this test group (expect the case 25). However, once we increase the $n_{presv}$ towards a large value, for instance, $n_{presv}=200$, EBR might provide a approximate models in a complex form. That is, the basis number of all regression models is larger than 20.

In this test group, the performance of EBR is also compared with the FFX's. Although the bases number of regression models of EBR is more than FFX's, its NMSEs is still much lower than FFX's, as shown in Table \ref{test_table_4}. The increasing of $n_{presv}$ will cause the increasing computation cost of EBR. So, in practical applications, we do not set $n_{presv}$ to a large value. $n_{presv} < 40$ is acceptable.

\begin{table*}
\scriptsize
\centering
\caption{Study on control parameter.}
\label{test_table_4}
\begin{tabular}{llllllll}
\hline
No. & Dim & Target model                       & Domain     & \begin{tabular}[c]{@{}l@{}}No.\\ bases\\ of EBR\end{tabular} & \begin{tabular}[c]{@{}l@{}}Test\\ error(\%)\\ of EBR\end{tabular} & \begin{tabular}[c]{@{}l@{}}No.\\ bases\\ of FFX\end{tabular} & \begin{tabular}[c]{@{}l@{}}Test\\ error(\%)\\ of FFX\end{tabular} \\ \hline
25  & 1   & $0.3x\sin \left( 2\pi x \right)$   & $[-3,3]$   & \textbf{21}                                                  & \textbf{1.70e-2}                                                  & 8                                                            & 19.74                                                             \\
26  & 1   & $\sin \left( x^3+x \right)$        & $[-3,3]$   & \textbf{22}                                                  & \textbf{2.93e-5}                                                  & 10                                                           & 29.61                                                             \\
27  & 1   & $\sin x \sin \left( x+x^2 \right)$ & $[-3,3]$   & \textbf{28}                                                  & \textbf{2.23e-20}                                                 & 9                                                            & 13.29                                                             \\
28  & 2   & $\sin x_1 +\sin x_2^2$             & $[-3,3]^2$ & \textbf{21}                                                  & \textbf{4.68e-8}                                                  & 16                                                           & 7.771                                                             \\ \hline
\end{tabular}
\end{table*}

\section{Conclusion}
\label{Section5}
A new deterministic algorithm, Elite Bases Regression (EBR), for symbolic regression has been proposed in this paper. It is a linear approximation method based on the generalized linear model (GLM). In EBR, all generated candidate bases are coded with parse-matrices in specific mapping rules. The correlation coefficients with respect to the target model are used to evaluate the candidate bases, and only a certain number of elite bases are preserved to form the regression model. This makes EBR easy to realize real-time computation. 

A comparative study between EBR and a recent proposed deterministic machine learning method for symbolic regression, Fast Function eXtraction (FFX), have been conducted. Numerical results indicate that EBR performs better for its more concise linear approximation regression models and lower normalized mean square error than FFX. Moreover, EBR can provide exact fitting models with regard to the target models, which shows the ability of \emph{structure optimization}.

As a future work, it is planned to study on improving the performance of EBR by introducing new modifications. For example, nonlinear correlation detection is desired, so that EBR can be applied to complicated real-world applications more effectively.

\section*{Acknowledgment}
This work has been supported by the National Natural Science Foundation of China (Grant No. 11532014).


\begin{thebibliography}{1}

\bibitem{Beachkofski2002}B. K. Beachkofski, R. V. Grandhi, Improved distributed hypercube sampling, in: Proceedings of  the 43rd AIAA/ASME/ASCE/AHS/ASC Structures, Structural Dynamics, and Materials Conference, Denver, Colorado, 2002.

\bibitem{Davidson2003}J.W. Davidson, D.A. Savic, G.A. Walters, Symbolic and numerical regression: experiments and applications, Information Sciences 150 (1) (2003) 95-117.

\bibitem{Deklel2016}A. K. Deklel, A. M. Hamdy, E. M. Saad, Multi-objective symbolic regression using long-term artificial neural network memory (LTANN-MEM) and neural symbolization algorithm (NSA), in: Neural Computing and Applications, Springer, 2016, pp. 1-8.

\bibitem{Deuze1989}J. L. Deuz\'{e}, M. Herman, R. Santer, Fourier series expansion of the transfer equation in the atmosphere-ocean system, Journal of Quantitative Spectroscopy \& Radiative Transfer 41 (6) (1989) 483-494.

\bibitem{Dong2015}X. Dong, W. Dong, Y. Yi, Y. Wang, X. Xu, The Recent Developments and Comparative Analysis of Neural Network and Evolutionary Algorithms for Solving Symbolic Regression, in: Intelligent Computing, Springer, 2015, pp. 703-714.

\bibitem{Ferreira2002}C. Ferreira, Gene expression programming in problem solving, Springer, 2002, pp. 635-653.

\bibitem{Gielen2013}G. Gielen, E. Maricau, Stochastic degradation modeling and simulation for analog integrated circuits in nanometer CMOS, in: Proceedings of Design, Automation \& Test in Europe Conference \& Exhibition (DATE), IEEE, 2013, pp. 326-331.

\bibitem{Goldberg1996}B. Goldberg, Functional programming languages, ACM Computing Surveys 28 (1) (1996) 249-251.

\bibitem{Holmes1996}P. Holmes, P. J. Barclay, Functional languages on linear chromosomes, in: Proceedings of the 1st annual conference on genetic programming, MIT Press, 1996, pp. 427-427.

\bibitem{Icke2013}I. Icke, J. C. Bongard, Improving genetic programming based symbolic regression using deterministic machine learning, in: Proceedings of the IEEE Congress on Evolutionary Computation, IEEE-CEC, 2013, pp. 1763-1770.

\bibitem{Dervis2012}D. Karaboga, C. Ozturk, N. Karaboga, B. Gorkemli,  Artificial bee colony programming for symbolic regression, Information Sciences 209 (2012) 1-15.

\bibitem{Koza1992}J.R. Koza, Genetic programming: on the programming of computers by means of natural selection, MIT Press, Cambridge, MA, USA, 1992.

\bibitem{Luo2015}C. Luo, Z, Hu, S. Zhang, Z. Jiang, Adaptive space transformation: An invariant based method for predicting aerodynamic coefficients of hypersonic vehicles, Engineering Applications of Artificial Intelligence 46 (2015) 93-103.

\bibitem{Luo2012}C. Luo, S. Zhang, Z. Jiang, Parse-matrix evolution for symbolic regression, Engineering Applications of Artificial Intelligence 25 (6) (2012) 1182-1193.

\bibitem{Maricau2012}E. Maricau, D. D. Jonghe, G. Gielen, Hierarchical analog circuit reliability analysis using multivariate nonlinear regression and active learning sample selection, in: Proceedings of Design, Automation \& Test in Europe Conference \& Exhibition (DATE), IEEE, 2012, pp. 745-750.

\bibitem{McConaghy2011}T. McConaghy, FFX: Fast, scalable, deterministic symbolic regression technology, in: Genetic Programming Theory and Practice IX, Springer, 2011, pp. 235-260.

\bibitem{Moller1997}T. Moller, R. Machiraju, K. Mueller, R. Yagel, Evaluation and design of filters using a Taylor series expansion, IEEE Transactions on Visualization and Computer Graphics 3 (2) (1997) 184-199.

\bibitem{Nelder1972}J. A. Nelder, R. W. M. Wedderburn, Generalized linear models, Journal of the Royal Statistical Society, 135 (1972) 370-384.

\bibitem{ONeil2003}M. O’Neil, C. Ryan, Grammatical evolution, IEEE Transactions on Evolutionary Computation 5 (2001) 349-358.

\bibitem{Patelli2010}A. Patelli, L, Ferariu, Elite based multiobjective genetic programming in nonlinear systems identification, Advances in Electrical and Computer Engineering 10 (1) (2010) 94-99.

\bibitem{Steinberg2006}D. M. Steinberg, D. K. J. Lin, A construction method for orthogonal Latin hypercube designs, Biometrika 93 (2) (2006) 279-288.

\bibitem{Worm2016}A. Worm, Prioritized Grammar Enumeration: A novel method for symbolic regression, Binghamton University - State University of New York, 2016 Ph.D. thesis.

\bibitem{Yang2005}Y. W. Yang, C.  Wang, C. K. Soh, Force identification of dynamic systems using genetic programming, International journal for numerical methods in engineering 63 (9) (2005) 1288-1312.

\end{thebibliography}
\end{document}